\documentclass{mn2e}
\usepackage{psfig}
\def    \kms    {\rm km\ s^{-1}}

\def    \modot  {\rm M_{\odot}}

\def\ec{{$\eta$ Carin\ae}}
\def\xmm{{XMM-Newton}}
\def\einstein{{EINSTEIN}}
\def\rosat{{ROSAT}}
\def\fo{{FO\,15}}
\def\pasp{{Publ. Astron. Soc. Pac.}}
\def\aap{{Astron. Astrophys.}}
\def\apj{{Astrophys. J.}}

\def\mnras{{Mon. Not. R. Astron. Soc.}}
\def\apjs{{Astrophys. J., Suppl. Ser.}}
\def\aj{{Astron. J.}}

\begin{document}
 
\title
[\fo\,: a new O-Type eclipsing binary in the Carina Nebula]
{Optical Spectroscopy of X-Mega targets in the Carina Nebula - VI. \fo\,:
a new O-Type double-lined eclipsing binary }  

\author[V. Niemela et al.]
{V.S. Niemela$^{1}$
\thanks{Member of Carrera del Investigador, CIC-BA, Argentina; Visiting Astronomer, CASLEO, San Juan, Argentina},
N.I. Morrell$^{2}$
\thanks{Member of Carrera del Investigador , CONICET, Argentina, on leave from 
La Plata National University, Argentina; Visiting Astronomer, CASLEO, 
San Juan, Argentina; Visiting Astronomer, CTIO, NOAO, operated by 
AURA, Inc., for NSF; Visiting Astronomer, European Southern Observatory, Chile},
E. Fern\'andez Laj\'us$^{1}$
\thanks{Fellow of CONICET, Argentina.},
\newauthor
R. Barb\'a$^{3}$
\thanks{Member of Carrera del Investigador, CONICET, Argentina, on leave
from La Plata National University, Argentina.},
J.F. Albacete Colombo$^{4}$
\thanks{Post-doctoral fellow of CONICET, Argentina.},
M. Orellana$^{5}$
\thanks{Fellow of CONICET, Argentina}
\\
$^{1}$ Facultad de Ciencias Astron\'omicas y Geof\'{\i}sicas de la Universidad
Nacional de La Plata, Paseo del Bosque S/N, 1900 La Plata, Argentina\\ 
$^2$ Las Campanas Observatory, Observatories of the Carnegie Institution of Washington, Casilla 601, La Serena, Chile\\
$^3$ Departamento de F\'{\i}sica, Universidad de La Serena, Benavente 980, La Serena, Chile\\ 
$^4$ Osservatorio Astronomico di Palermo, Piazza del Parlamento 1, Palermo (90141), Italy\\ 
$^5$ Instituto Argentino de Radioastronom\'{\i}a, C.C.~5, 1894 Villa Elisa, Argentina\\ 
}                                                                           

\date{}
\pagerange{\pageref{firstpage}--\pageref{lastpage}}
\pubyear{2005}

\label{firstpage}

\maketitle

\begin{abstract}        
We report the discovery
of a new O--type double--lined spectroscopic binary with a short orbital period
of 1.4 days. We find the primary component of this binary, \fo\,, to have 
an approximate spectral type O5.5Vz, i.e. a Zero--Age--Main--Sequence star.
 The secondary appears to be of spectral type O9.5V. 
We have performed a numerical model fit to the public ASAS photometry, 
which shows that \fo\ is also an eclipsing binary. We find an orbital 
inclination of $\sim 80^{\circ}$.  
From a simultaneous light-curve and radial velocity solution we find the masses 
and radii of the two components to be $30\pm1$ and $16\pm1$ solar masses and
$7.5 \pm 0.5$ and $5.3 \pm 0.5$ solar radii. These radii, and hence also the 
luminosities, are smaller than those of normal O-type stars, but similar to 
recently born ZAMS O-type stars. The absolute magnitudes derived from our 
analysis locate \fo\, at the same distance as \ec\,. From Chandra and XMM 
X-ray images
we also find that there are two close X-ray sources, one coincident with \fo\,
and another one without optical counterpart. This latter seems to be a highly 
variable source, presumably due to a pre--main--sequence stellar neighbour 
of \fo\,. 

\end{abstract}

\begin{keywords}
stars: binaries, spectroscopic, eclipsing
stars: O--type -
stars: fundamental parameters -
stars: individual: \mbox \fo\,
- X-rays: stars
\end{keywords}
 
\section{Introduction}                                                       
In their survey for OB stars in the field of the Carina Nebula (NGC\,3372),
Forte \& Orsatti (1981) discovered an early O type star in the
darkest region of the nebula. This star 
($\alpha_{2000} = 10^{\rm h}45^{\rm m}36^{\rm s}$;
$\delta_{2000} = -59^{\circ}48'22''$; V\,=\,12.05), number 15 in their list
of new OB stars, was assigned a spectral classification O4V on photographic
image tube spectrograms. The star has been named \fo\, in subsequent 
literature. 

In an infrared study  of the stellar population in the direction
of the Carina Nebula, Smith (1987) identifies \fo\, as a member of a group of
heavily reddened OB stars in the southeast border of the open cluster Trumpler
16, 
considering that these stars most probably also are members of
this cluster. An anomalous reddening law characterized by a value of
R=Av/E(B-V)=4.8 is obtained by Smith (1987).

\begin{figure*}
\centering
\vspace{15cm}
\includegraphics{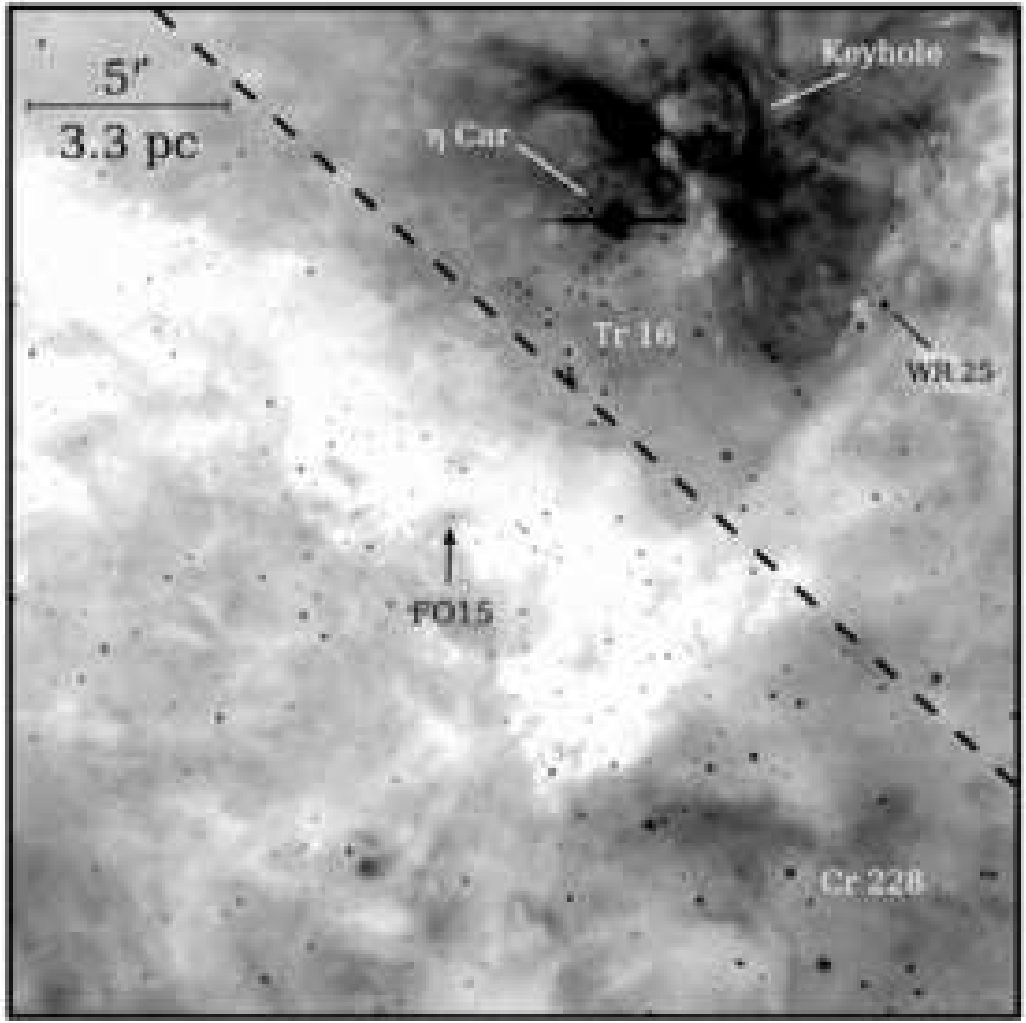}
\caption{ H$\alpha$ Image of the Carina Nebula field surrounding \fo\ 
(indicated by an arrow). North is up and East to the left. 
The dashed line corresponds to the NW limit of the Spitzer Space Telescope IR
observation by Smith et al. 2005 (see text).}
\label{mapa}
\end{figure*}

In Fig.~\ref{mapa} we illustrate the location of \fo\ in the 
Carina Nebula, inside
the V--shaped dust lane dividing the brightest part of the nebula,
between the open clusters Trumpler 16 and Collinder 228. The H$\alpha$ 
image shown in Fig.~\ref{mapa} was obtained in 1999, May, 
with the Curtis-Schmidt telescope at the Cerro Tololo Inter--American 
Observatory (CTIO), Chile.

FO15 was observed as a faint X-ray source in Einstein satellite X-ray images 
of the Carina Nebula (Chlebowski et al. 1989). Because one of the most often
proposed mechanisms for producing X-rays from O-type stars are colliding
stellar winds in binary systems, we decided to include FO15 in our
ongoing optical spectroscopic observations in search for O-type binaries.

In this paper, we will present our spectroscopic observations of FO15
showing this star to be a double--lined binary system with a short orbital
period. After our spectroscopic analysis was essentially complete, FO15
also appeared as an eclipsing binary in the All Sky Automated Survey (ASAS)
(cf. Pojma\'nski, 2003).
We have analyzed the ASAS light curve together with our radial velocity
study of the binary orbit.

\section{Optical spectroscopy} 

Twenty five
spectral images of FO15 were obtained with the 2.15-m telescope at
Complejo Astron\'omico El Leoncito (CASLEO)\footnote{CASLEO is operated under
agreement between CONICET, SeCyT, and the National Universities of La Plata,
C\'ordoba and San Juan, Argentina} in San Juan, Argentina, during several
observing runs between February 1996 and March 2003.   
We used, alternatively, the Cassegrain
Boller \& Chivens (B\&C) spectrograph, with a 600 l\,mm$^{-1}$ grating and a
PM~516$\times$516  20$\mu$m pixel CCD as detector, and the REOSC echelle
spectrograph, in its
Simple Dispersion mode.   For the REOSC spectra a TEK~1024$\times$1024 pixel
CCD,  with pixel size of 24$\mu$m was used as detector.  These instrumental
configurations provide  reciprocal dispersions of $\sim$2.3  and
1.6 \AA\,px$^{-1}$, respectively. The wavelength region observed was
roughly $\lambda\lambda$3900 -- 5040~\AA.
 
We used a slit width of 2 arcsec for all our CASLEO spectra.
Exposure times for the stellar images ranged between 30 minutes and 1 hour,
resulting in spectra of signal-to-noise ratio between 50 and 150,
depending on seeing and transparency.
                                 He-Ar and Cu-Ar comparison arc images
were observed  with the B\&C and REOSC spectrographs,
respectively, at the same telescope position
as the stellar images immediately after or before the stellar exposures.
The usual series of  bias and flat-field frames were also obtained
for each observing night.
    
One high resolution spectral image of \fo\,
was obtained with  ESO-VLT UT2 (Kueyen) and the UV-Visual
Echelle Spectrograph (UVES) in December 2001.
The setting used produced a blue spectrum covering the region 3260-4520\AA \
              with a resolution (2 pixels) R = 67000 and a red one
spanning from 4580 to 6690\AA\,
at a spectral resolution (2 pixels) R = 58000.
High gain, 2 $\times$ 2 binning mode was used.
A signal-to-noise of $\sim$50 was achieved in this observation.
     
Further three high resolution spectra were obtained at Las Campanas
Observatory (LCO), as follows:
one in December 2002, with the Magellan II (Clay)
6.5-m telescope and Magellan Inamori Kyocera Echelle (MIKE)
double echelle spectrograph (Bernstein et al. 2003).  
A 1 $\times$ 3 binning was applied achieving a
resolution of 19000 in the red spectrum (4500 - 7200 \AA)
and 29000 in the blue spectrum (3200-4700 \AA), at a signal-to-noise
ranging from 60 to 200 in the red and from 40 to 100 in the blue
in a 1200\,s exposure using a 1 arcsec slit width.
Another echelle observation was obtained
in May 2003 with the Magellan I (Baade) 6.5-m telescope
and MIKE. No binning was applied to these data yielding a
resolution $R \sim 36000$ in the blue spectrum
(3200 - 4800\AA) and 27000 in the red spectrum (4800 - 8800\AA),
with a 0.7 arcsec slit width.
The signal-to-noise achieved in this 600 seconds exposure is
60-130 in the red spectrum and 10-60 in the blue spectrum.
A third echelle spectrum of \fo\ was obtained with the
2.5m du Pont telescope and its echelle spectrograph in July 2003.
The signal-to-noise achieved in the 3 $\times$ 1200 seconds exposure
ranges from 10 to 140 from the blue (3700 \AA) to the red end (10000 \AA)   
of the spectrum at a resolution $R \sim 23000$ using a 0.75 arcsec slit. 

Th-Ar comparison lamps were secured for every high
resolution observation as well as the usual series of
bias and flat-field exposures.     

All the spectral images were processed and analyzed with
standard IRAF routines. MIKE spectra were extracted using the
IRAF tasks contained in the {\it mtools} package,
developed by J. Baldwin and available for
downloading from LCO website.
                              
\section {The spectrum of \fo\,}
\begin{figure*}
\centering
\vspace {12cm}
\includegraphics{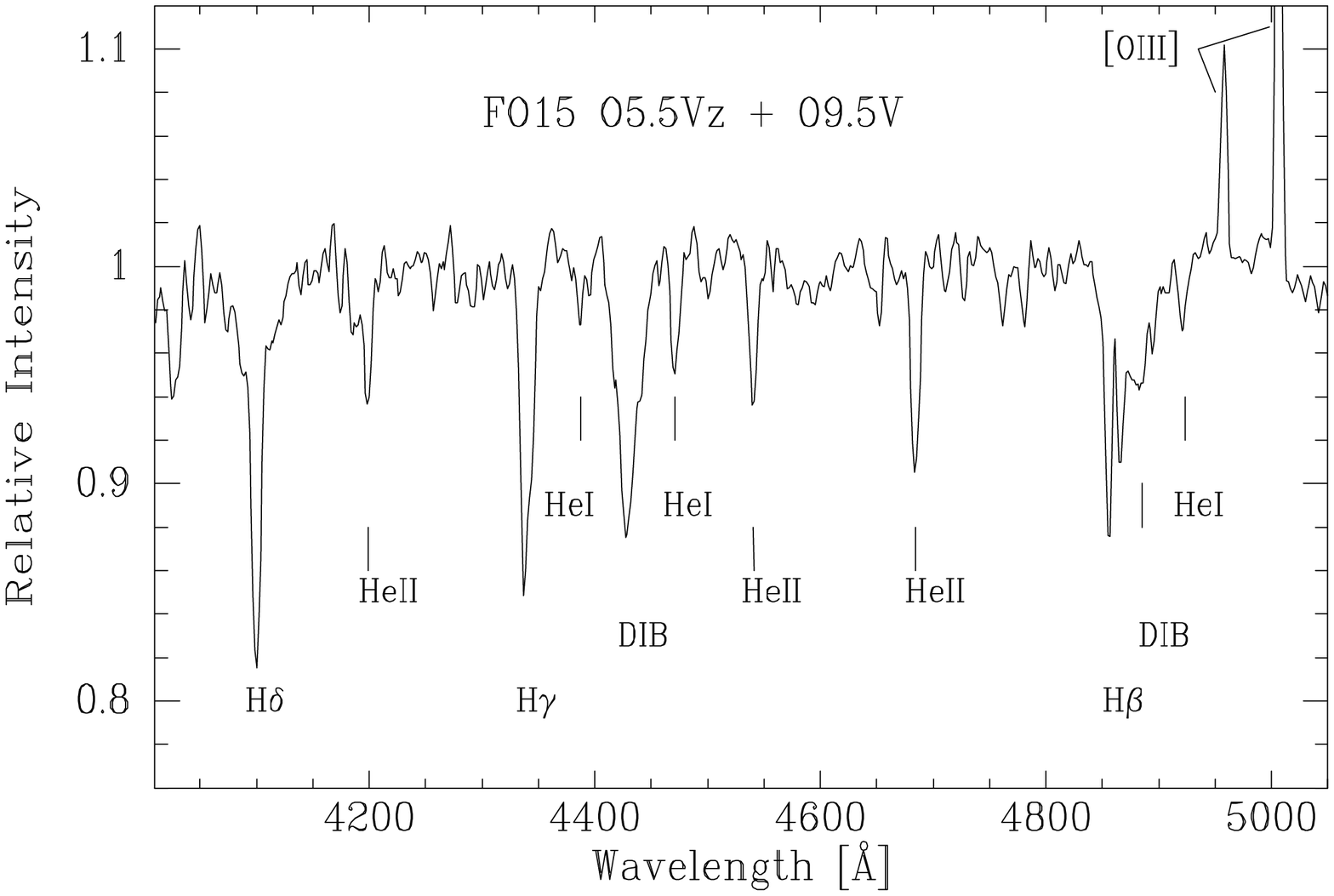}
\caption {Continuum rectified spectrum of \fo\ obtained at CASLEO in 2001,
February. The spectrum corresponds to the binary phase 0.43, with the primary
component in front of the system. The spectral features identified are the 
stellar absorptions of 
hydrogen H$\delta$, H$\gamma$, H$\beta$; of He{\sc i} $\lambda\lambda$4387, 
4471, 4921; and of He{\sc ii}  $\lambda\lambda$4200, 4541, 4686. Many
interstellar absorptions and nebular emission are also observed in this
spectrum.  }
\label{fig02}
\end{figure*}

Our spectra of \fo\, confirm that the stellar spectrum corresponds to 
an O--type star, as illustrated in Fig.\ref{fig02}. In this figure 
we also observe 
that the He{\sc ii} absorption at $\lambda$4686 is stronger than any of the 
other He{\sc ii} lines, a proposed signature of a Zero--Age--Main--Sequence 
luminosity class (Vz) for O stars (cf. Walborn \& Blades 1997 and references 
therein). 

\begin{figure*}
\centering
\vspace {12cm}
\includegraphics{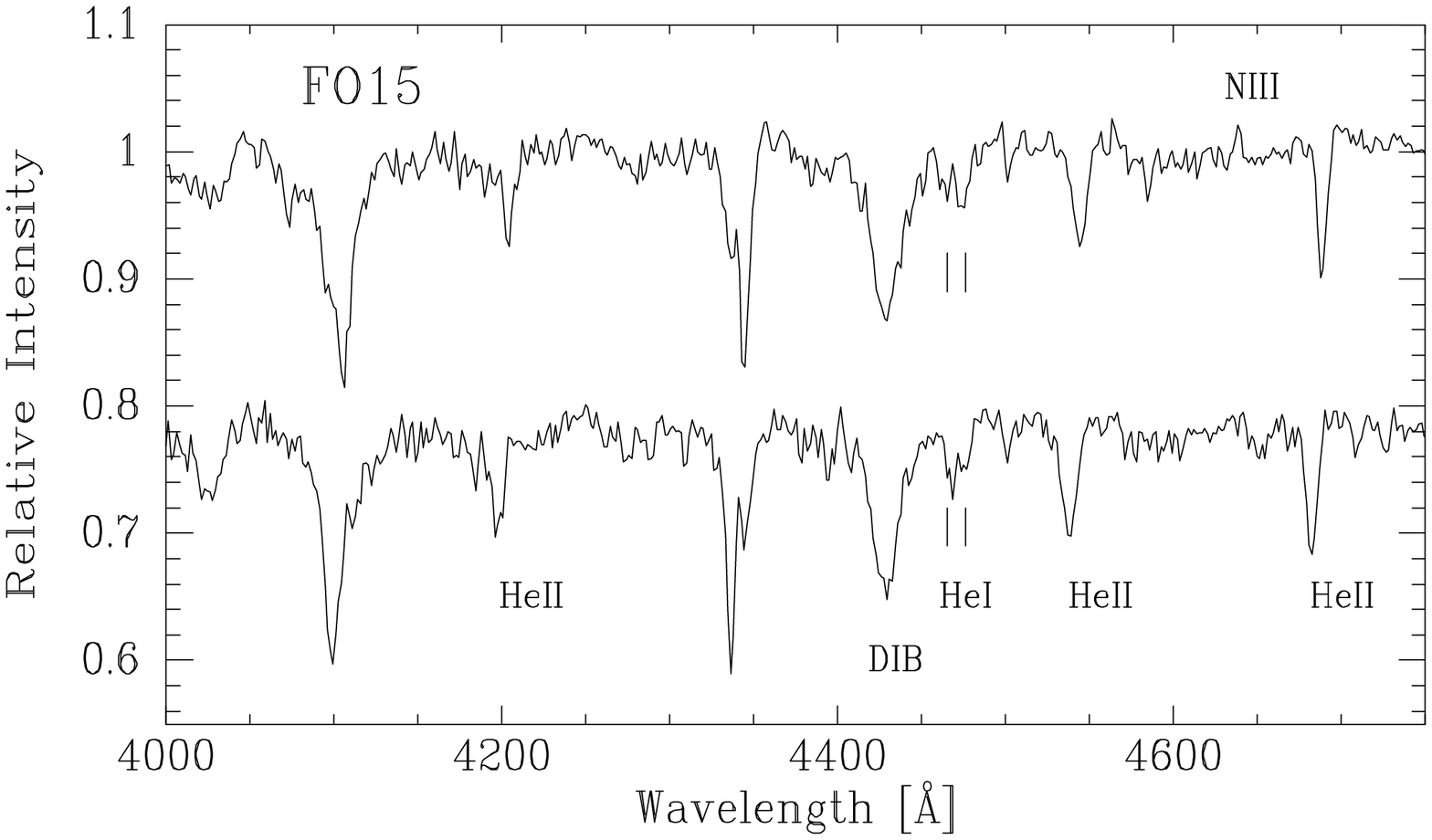}
\caption { Spectra of \fo\, obtained at CASLEO in 2001,February, during 
orbital phases 0.78 (upper) and 0.14 (lower) showing the
spectral lines of the two components seen in He{\sc i} $\lambda$ 4471.
Note also the faint emission line of N{\sc iii}.}
\label{fig03}
\end{figure*}

A visual inspection of 
our first spectra of \fo\, which we obtained at CASLEO in 1996, December, 
during successive nights, showed that the stellar Hydrogen absorption lines 
moved from the blueside to the redside of the nebular emissions, a 
signature of a rather high amplitude orbital motion in a binary system. 
This was confirmed with further observations, which also showed that the 
neutral Helium lines appeared double at maximum velocities, but ionized  
Helium lines always appeared single, as shown in Fig.~\ref{fig03}, 
which depicts 
two spectra obtained at CASLEO during approximately opposite orbital phases 
of the binary (see below). 

The spectrum of \fo\, appears somewhat variable, as occasionally faint 
N{\sc iii} emisson at $\lambda$ 4634-40 appears, as seen in the upper 
spectrum of Fig.\ref{fig03}. However, the radial velocities of this emission 
do not clearly correspond to any of the binary components, and it may arise
in the zone of interaction of the close components.

Nebular emission lines are observed in our spectra, as well as nebular
He{\sc i} $\lambda$ 3888 absorption. The interstellar absorption lines of 
Ca{\sc ii} and
Na{\sc i} appear multiple. These are common features observed towards the
giant H{\sc ii} region of Carina (e.g. Walborn \& Hesser 1975, Walborn et al.
2002). 
The main components of Ca{\sc ii} absorption in our echelle spectra 
have a heliocentric radial velocity of $+7\pm1~\kms$, the nebular [O{\sc iii}]
 $\lambda$ 5007 emission a velocity of $-19\pm2~\kms$, and the nebular
He{\sc i} $\lambda$ 3888 absorption of $-22\pm2~\kms$. We have used these
velocities to check the consistency of the stellar radial velocities 
derived from our lower resolution spectra.
     
\subsection {Spectral types of the binary components}
Spectral classification of components in a close binary systems is
not as straightforward task as classifying spectra of single stars.
In very close systems of hot stars, such as FO15,
mutual heating effects may introduce a spectral appearance with hotter
effective temperature. Furthermore, spectral variations due to the
interaction of stellar winds of the components are often observed.

We have chosen the usual approach to classify the binary components of
\fo\, in the spectrograms observed when the spectral lines have their
maximum separation.
To determine the spectral types, we measured in our higher resolution 
spectra the
equivalent widths of the He{\sc i} and He{\sc ii} lines. We then used the
cuantitative classification criteria for O-type stars as described by 
Conti \& Alschuler (1971), comparing the equivalent
widths of He{\sc i}$\lambda$\,4471 vs. He{\sc ii}$\lambda$\,4542. 
We also compared
 He{\sc i}$\lambda$\,4922 vs.  He{\sc ii}$\lambda$\,5411, as suggested 
by  Kerton et al. (1999).
According to these criteria, we obtained for the primary component of \fo\
a spectral classifications of O5.5V.

We note however, that the spectrum observed near the orbital phase when
the primary star is in front of the system (see Fig.~\ref{fig02}) and the 
contribution of the
secondary to the spectrum is expected to be minor, corresponds to a spectral
type about one subtype later than O5.5, when compared with the digital 
spectral atlas of O-type stars published by Walborn \& Fitzpatrick (1990).

\begin{figure*}
\vspace {13cm}
\includegraphics{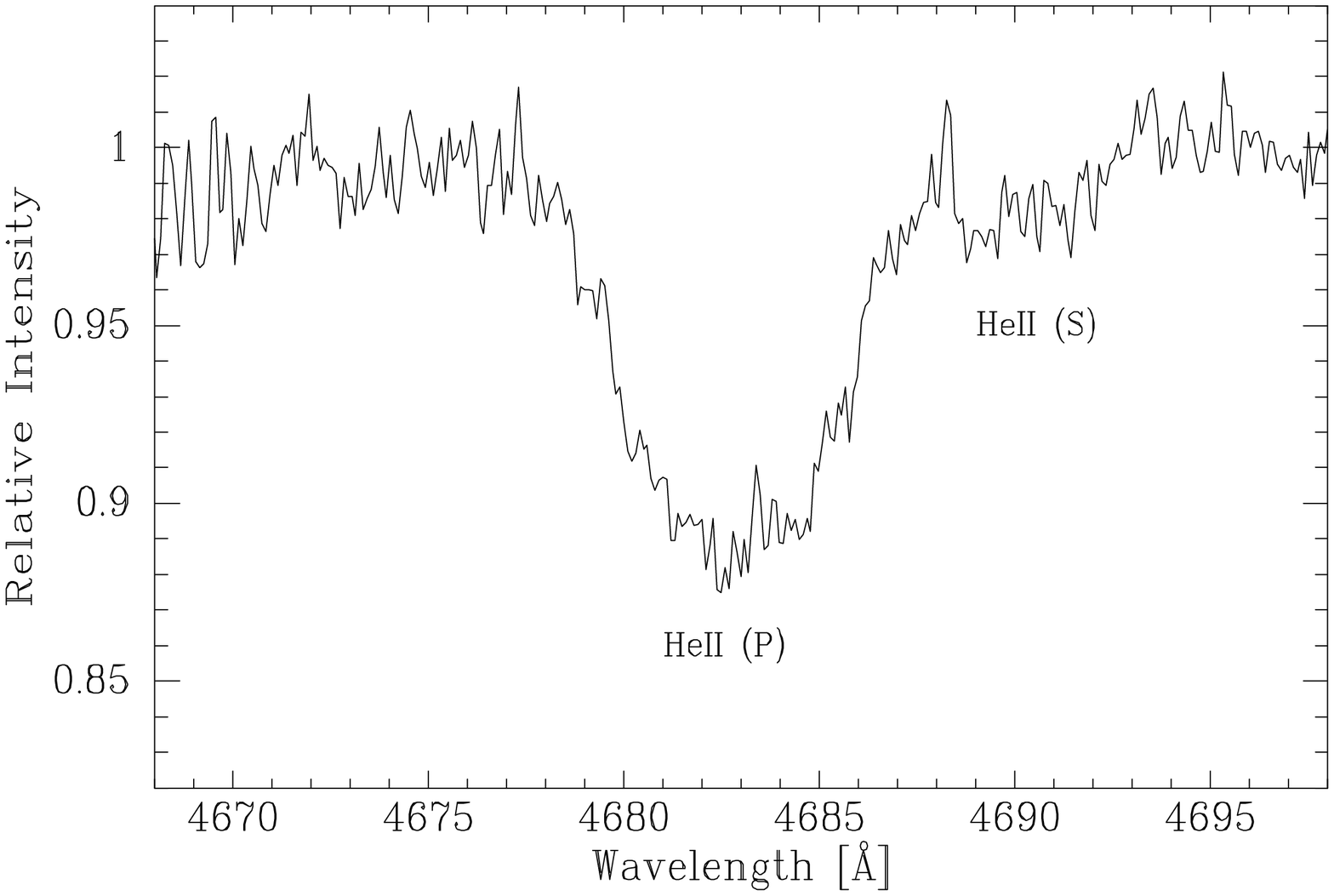}
\caption { Continuum normalized spectrum of \fo\,, obtained with MIKE at
the Magellan II telescope in 2002, December, (orbital phase 0.34) showing the
He{\sc ii} $\lambda$ 4686 absorption lines of both binary components.}
\label{fighe2}
\end{figure*}

The spectral type of the secondary component is more difficult to ascertain.
He{\sc ii} absorptions correponding to the secondary component is not observed in our
lower resolution spectra, as is evident in Fig.~\ref{fig03}.
In our high  resolution spectra all the He{\sc i} absorptions appear
as double lines, but only
the He{\sc ii}$\lambda$\,4686 line corresponding to the secondary component
is clearly observed. Both components of this
line are shown in Fig.~\ref{fighe2}. He{\sc ii}$\lambda$\,5411 absorption of
the secondary component may also be present, but it appears blended with the
diffuse interstellar bands at $\lambda\lambda$\,5404 and 5420. The spectral 
type of the secondary component probably is O9.5V, but could be slightly 
later. 

\begin{figure*}
\vspace {13cm}
\includegraphics{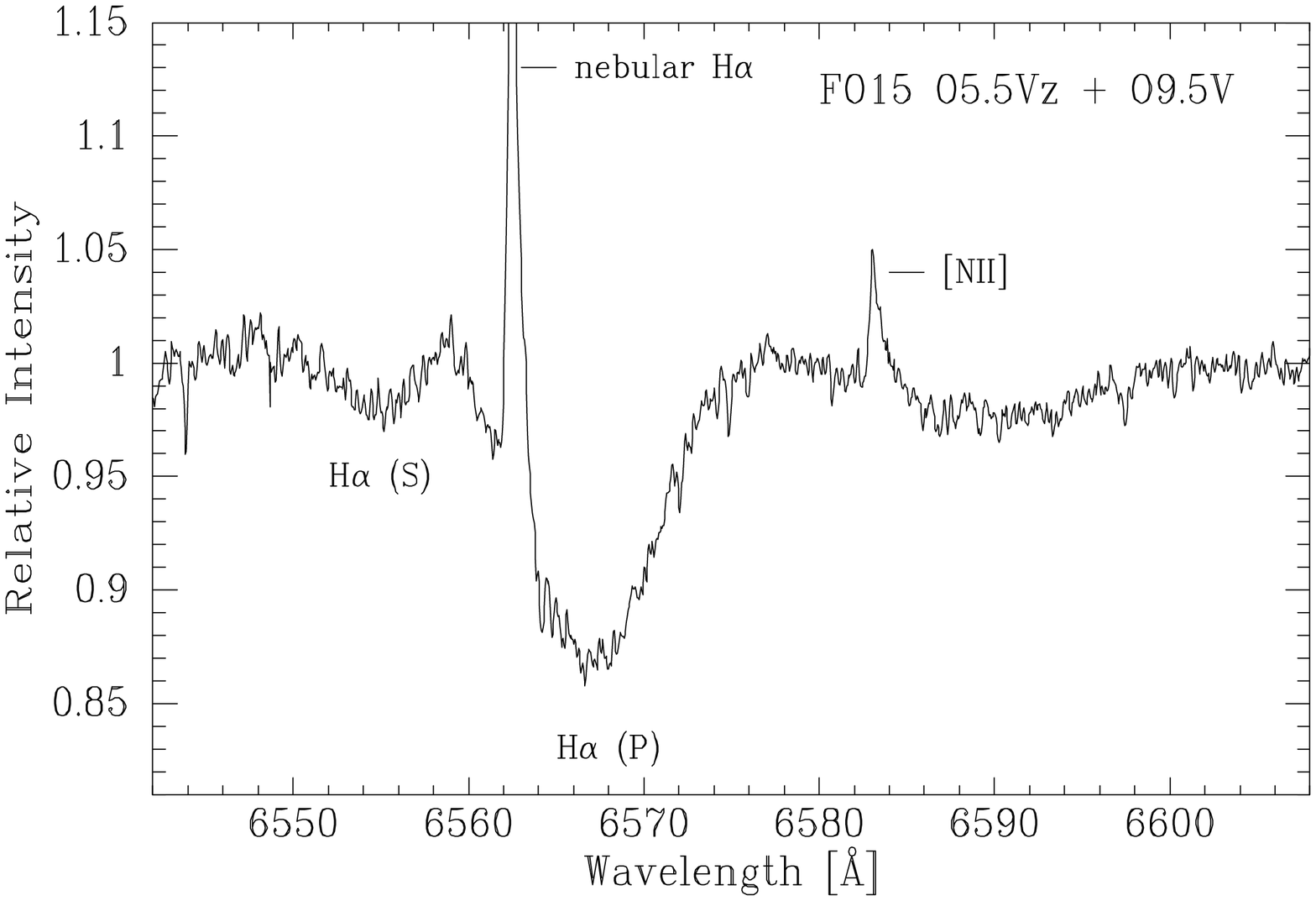}
\caption { Continuum normalized spectrum of \fo\,, obtained with MIKE at
the Magellan I telescope in 2003, May, (orbital phase 0.81) showing the
H$\alpha$ $\lambda$ 6563 absorption lines of both binary components. Nebular
emissions of H$\alpha$ and [N{\sc ii}] are also indicated in the spectrum.}
\label{figha}
\end{figure*}

We do not observe in our optical spectra of \fo\ any mass-loss
indicators (signs of strong stellar winds), such as those listed
e.g. by Hutchings (1978). The Balmer H$\alpha$ line is observed in
absorption in both binary components,  as illustrated in Fig.~\ref{figha}. 
The radial velocity of this absorption agrees with the values derived from 
He lines, indicating that the H$\alpha$ absorption is mostly photospheric, and
not formed in the accelerated part of an expanding atmosphere.
Thus \fo\ may be composed of weak-wind and low mass-loss rate
O type dwarfs, as those recently discussed by Martins et al. (2005b).

\section {The radial velocity orbit}
To determine the radial velocity orbit of the binary,
we measured the radial velocities of the He lines in the spectra
of \fo\, fitting gaussian profiles to the spectral lines within the
IRAF routine SPLOT. The results are listed in Table \ref{vr}.
 The radial velocities of the primary component are mostly based on the
velocities of He{\sc ii} absorptions, and the velocities of the
secondary on those of He{\sc i}.
To calculate the orbital parameters, we used a modified version of the 
program originally written by 
Bertiau \& Grobben (1976), introducing the period determined by ASAS
photometry, namely 1.4136 days, as an initial value. We assigned higher 
weight to the higher resolution observations and to those obtained near 
quadratures.
The orbital parameters are listed in Table~2, along with their formal
 standard errors as calculated by the above mentioned program.

\begin{table}
\caption{Observed heliocentric radial velocities for the primary and secondary
components of \fo\,.
Radial velocities and (O-C) values are in [km\,s$^{-1}$].}
\label{vr}
\begin {tabular}{lclllll}
\hline
HJD (days)&Phase&\multicolumn{2}{c}{Primary}&&\multicolumn{2}{c}{Secondary}\\
\cline{3-4} \cline{6-7}
2400000+  &$\phi$&RV  & O-C && RV & O-C \\
\hline
50125.751(a)  &0.57& 102  &-7   &&    &     \\
50127.747(a)  &0.98&  -4  & 14   &&    &     \\
50128.758(a)  &0.70& 222  & 10   &&-423&  25 \\
50129.694(a)  &0.36&-165  &  9   &&    &     \\
& && & \\
50471.804(a)  &0.38&-180  &-28   &&    &     \\
50472.773(a)  &0.07&-148  & -15  &&    &     \\
50473.778(a)  &0.78& 204  & -3   &&-431&  8 \\
50477.807(a)  &0.63& 167  & -1   &&    &     \\
50478.774(a)  &0.31&-233  &-15   &&    &     \\
& && & \\
50854.825(a)  &0.34&-211  &-18   &&    &     \\
50858.785(a)  &0.14&-218  & -6    && 367& 6 \\
50860.749(a)  &0.53&  15  &-45   &&    &     \\
50861.739(a)  &0.23&-247  & -1   && 426&  -1 \\
& && & \\
51355.466(a)  &0.51&  18  &-11   &&    &     \\
& && & \\
51653.594(a)  &0.42& -93  & 13   &&    &     \\
51654.551(a)  &0.10&-173  & -7   &&    &     \\
& && & \\
51718.487(a)  &0.33&-206  & 2    && 361&  8  \\
& && & \\
51959.799(a)  &0.04&-109  &-15   &&    &     \\
51960.842(a)  &0.78& 199  &-9    &&-452& -11 \\
51961.765(a)  &0.43& -76  & 16   &&    &     \\
51962.776(a)  &0.14&-219  & -8   && 369&  9  \\
& && & \\
52251.847(b)  &0.64& 179  &  -1  &&-382&   5 \\
& && & \\
52298.849(a)  &0.89& 106  & -3   &&    &     \\
52302.856(a)  &0.73& 202  & -14  &&-439&  18 \\
& && & \\
52329.604(a)  &0.65& 196  & 10   &&-386&  12 \\
52332.820(a)  &0.93&  80  & 12   &&    &     \\
& && & \\
52628.844(c)  &0.34&-192  & 2    && 345&  18 \\
& && & \\
52690.606(a)  &0.04&-113  & -24  &&    &     \\
& && & \\
52766.626(c)  &0.81& 200  &  12  &&-399&  2 \\
& && & \\
52836.495(d)  &0.24&-225  &  25  && 420&   -6 \\
\hline
\end{tabular}

\medskip
Notes:\\
Data origin: (a) CASLEO, (b) ESO-VLT, (c) LCO-Magellan, (d) LCO-du Pont\\
Phases have been calculated according to the ephemeris\\
HJD($\phi=0$) = 2452837.565 + 1.41356E
\end{table}

Our radial velocities define a circular orbit within the errors, 
the calculated orbital eccentricity being e=0.05$\pm$0.06.
This is as expected for a massive binary with such a short orbital period,
as is the case for \fo\,.

\begin{table}
\caption{ Parameters of Circular Radial Velocity Orbit for \fo\,}
\begin{center}
\begin{tabular}{ccc}
\hline
         & Primary & Secondary\\
\hline
&&\\
$P$ [days] & \multicolumn{2}{c}{1.41356$\pm$0.000003}\\
$V_{o}$[km s$^{-1}$]    & \multicolumn{2}{c}{-15$\pm$2}  \\

$a \sin i$ [R$_\odot$]  & 6.42$\pm$0.05  & 12$\pm$0.05  \\
$K$ [km s$^{-1}$]       &  231$\pm$2 & 442$\pm$3 \\
$M \sin^3i$ [M$_\odot$] & 29$\pm$1   & 15$\pm$1   \\
$T_{o}$ [HJD] 2.450.000+&3159.5$\pm$0.2&3150.7$\pm$0.2\\
\hline
\end{tabular}
\end{center}
\label{tab02}
\end{table}                                                                     

\section {Light curve analysis}

Photometric V filter data of FO15 were reported by The All Sky Automated 
Survey (ASAS) in Pojma\'nski (2003).
This star is catalogued as ASAS 104536-5948.4 in the ASAS Catalog of 
Variable Stars,
beeing classified as an eclipsing binary  with a
period of P=1$.\!{\!^d}$4136. 
A visual inspection of the ASAS light curve of FO15 shows 
periodic light variations with a rather 
large apparent scatter of data, almost $\pm0.1~mag$ over all orbital phases.
In order to obtain a first estimation of the orbital inclination $i$ of the 
binary system, we have attempted to fit a numerical eclipsing binary 
model to the ASAS observations, using the
Wilson-Devinney (W-D) Code (Wilson \& De\-vinney 1971, Wilson 1990, 
Wilson \& Van Hamme 2004).
ASAS photometric data and our radial velocities were used to 
compare with the results of the model fitting.

\begin{figure*}
{\hspace*{-2.0cm}
\psfig{figure=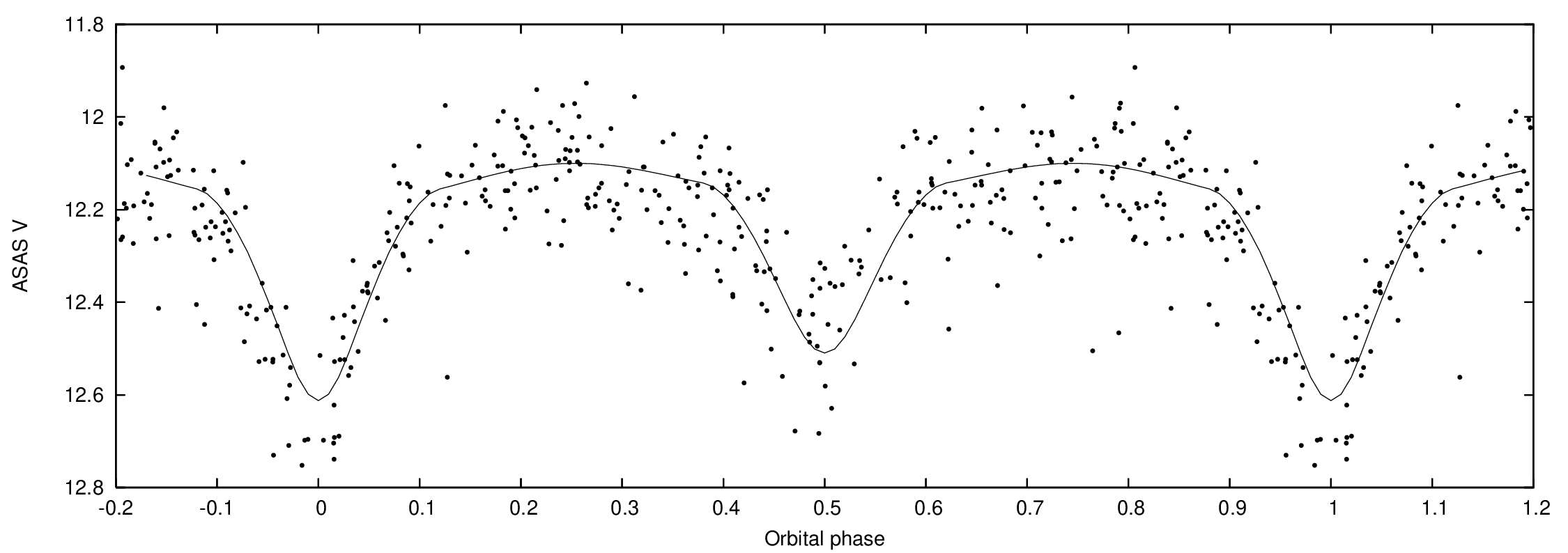,width=15.cm,angle=-90}}
\caption{ASAS V light curve of FO15. The continuous line represents our best
fit W-D model.
}
\label{CL}
\end{figure*}

\begin{figure*}
\centering
{\hspace*{-2.0cm}  \psfig{figure=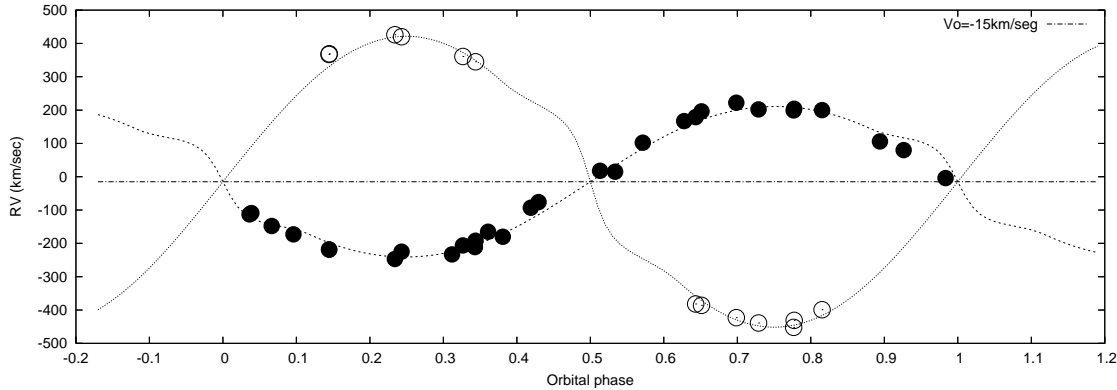,width=15.cm,clip=,angle=-90}}
\caption{Observed radial velocities of the primary (filled circles) and
secondary (open circles) of \fo\, as a function of the orbital phase.
The continuous curves represent the radial velocity orbit obtained from 
the W-D model fitting.
}
\label{VR}
\end{figure*}

We set the W--D code in Mode 2 for detached binaries with no constraints 
on the potentials (except the luminosity of the secondary).
The simplest considerations were applied for the emission parameters of 
the stars in the model, i.e. the stars as black bodies, approximate 
reflection model (MREF=1). No third light or spots were included. 
We adopted gravity darkening exponents $g_{1} = g_{2} = 1$, and 
bolometric albedos $Alb_{1} = Alb_{2} = 1$ were set for radiative envelopes.
We used the square root limb darkening law. Limb darkening coefficients 
for visual wavelengths were taken from D\'{\i}az-Cordov\'es, Claret \& 
Gim\'enez (1995), and bolometric
limb darkening coefficients from Van Hamme (1993).
We adopted the period P = 1$.\!{\!^d}$41356 and
mass ratio $M_{2}/M_{1} = 0.52$ from the radial velocity orbit (cf. Table~2). 
For both binary components, effective temperatures corresponding to their 
spectral types were adopted from the compilation by
Martins et al. (2005a), i.e. $T_{eff_{1}} \sim 40000 K$ 
and $T_{eff_{2}} \sim 32000 K$ for the primary and secondary components,
respectively. 
We assumed that the system has a circular orbit ($e = 0$) with both 
components rotating synchronously ($F_{1} = F_{2} = 1$), as suggested by 
the observations and as expected for a massive binary with a short period, 
as is the case for \fo\,. 
With all of these parameters fixed, we generated synthetic light and 
radial velocity curves adjusting them to the observations. 
We used the PHOEBE package (Pr\v sa \& Zwitter 2005) which incorporates 
numerical innovations and technical aspects to the W--D code.
The results for the best fit, optimised by the code, are shown in 
Figures~\ref{CL} and \ref{VR} and listed in Table~3.

\begin{table}
\center
\caption{Astrophysical data for the binary components of \fo\ derived from the
best fitting W-D model. $R_{L}$ stand for the effective radius of the Roche lobe}
\label{t2}
\begin{tabular}{c c c}
\noalign{\smallskip}\hline\noalign{\smallskip}
Parameters  &\multicolumn{2}{c}{Component}\\
            &    Prim.    &      Sec.     \\
\noalign{\smallskip}\hline\noalign{\smallskip}
P ($days$)     &\multicolumn{2}{c}{$1.41356 \pm 10^{-5}$}\\
$i$ ($^\circ$)      & \multicolumn{2}{c}{80 $\pm$ 1}\\
a ($R_{\odot}$)     &\multicolumn{2}{c}{19 $\pm$ 0.3}\\
M ($M_{\odot}$)     & 30.4 $\pm$ 1  & 15.8 $\pm$ 1\\
$M_{2}$/$M_{1}$     &\multicolumn{2}{c}{0.52$^\ast$}\\
R mean ($R_{\odot}$)     & 7.5 $\pm$ 0.5 &  5.3 $\pm$ 0.5 \\
$R_{L}$ ($R_{\odot}$)   & 8.3  &  6.15  \\
Teff ($^\circ K$)$^\ast$ &   40000   &      32000\\
Mbol                &-7.98$\pm$0.02 & -6.27$\pm$0.02 \\
$L_{2}/L_{1}$   &\multicolumn{2}{c}{0.35 $\pm$ 0.01}\\
Log g (cgs)         &4.17$\pm$0.01  & 4.19$\pm$0.05 \\
\noalign{\smallskip}\hline\noalign{\smallskip}
\multicolumn{3}{l}{$\ast$ :Fixed}
\end{tabular}
\end{table}

With the inclination of the orbital plane, which we found to be
$i = 80 \pm 1 ^{\circ}$, 
the values of the stellar masses for the components of \fo\,
$M_{1} = 30.4 \pm 1 M_{\odot}$ and
$M_{2} = 15.8 \pm 1 M_{\odot}$,
are in fair agreement with the tabulations by Martins et al. (2005a) based
on models of stellar atmospheres. 

\begin{figure*}
\vspace {10cm}
\includegraphics{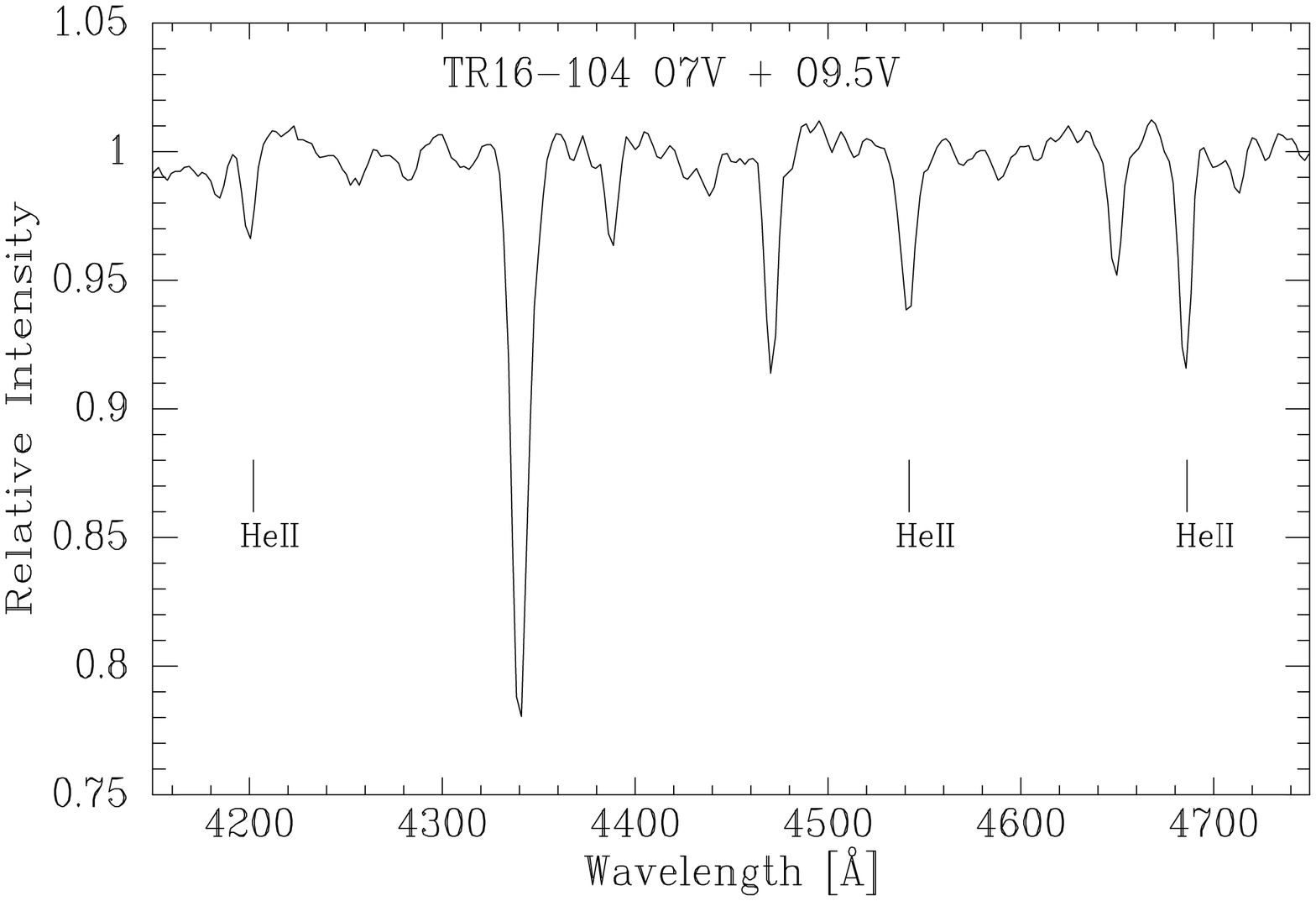}
\caption {Continuum normalized spectrum of Tr16-104 obtained at CASLEO in 1998,
February, showing the signature of Vz characteristic. 
Absorption lines of He{\sc ii} $\lambda\lambda$ 4200, 4541 and 4686 are identified. 
}
\label{sp104}
\end{figure*}

However, the stellar mean radii that we have obtained, namely
$R_{1} = 7.5 \pm 0.5 R_{\odot}$ and
$R_{2} = 5.3 \pm 0.5 R_{\odot}$
for the primary and secondary components, respectively,
are about 30\% smaller than the tabulated values, hence the bolometric 
luminosities implied from our light--curve solution are also lower. 
\footnote {A lower
$T_{eff}$ for the secondary component, would it be e.g. of spectral type B0V,
does not modify this result beyond the quoted error bars.}
This is similar to what was observed for another O--type binary system in the
Carina nebula, namely Tr16-104 by Rauw et al. (2001), who interpreted the
observations as due to fainter absolute magnitudes for stars just 
entering to the main sequence, i.e. stars of luminosity class Vz. In fact the
spectrum of Tr16-104 also exhibits the spectral signature of ZAMS O-type
stars, namely that He{\sc ii}$\lambda$4686 absorption is stronger than other
He{\sc ii} lines, as is illustrated in Fig.~\ref{sp104}.

The case may apply also to \fo\, but for a more reliable
assesment of this problem, an improved and less noisy light curve is needed, 
and is currently being obtained (Fern\'andez Laj\'us et al. 2006, 
in preparation).

\section{\fo\, in the Carina Nebula }
Based on their proximity in the sky, Forte \& Orsatti (1981) assumed that 
\fo\ belongs to the open cluster Tr 16, of which $\eta$ Car is the 
brightest member. However,
the line of sight in this direction of the Galaxy goes almost parallel to the 
 Sagittarius--Carina spiral arm, which is rich in young stellar population.
An O5V type star with the colors and apparent magnitude as those observed for
\fo\ would have a spectrophotometric distance of $\sim$5~kpc, if normal interstellar
extinction is assumed. This would place \fo\ well behind the Carina Nebula
and clusters embedded within. On the other hand, it is well known that the 
total--to--selective
extinction ratio (R) in the direction of Carina Nebula is anomalous (e.g.
Smith 1987, Tapia et al. 1988). Therefore,
we decided to determine the value of R for \fo\ using the published UBV
photometry from Forte \& Orsatti (1981) and IR magnitudes from the 2MASS
All-Sky Catalog of Point Sources (Cutri et al. 2003).

We used the code CHORIZOS developed by Ma\'{\i}z-Apell\'aniz (2004)
to derive the value of R for \fo . CHORIZOS is a code 
that uses $\chi^2$ minimization to find all models of energy distribution
compatible with an observed data set in the $N$-dimensional model
parameter space, which in our case are broadband photometry and spectral type.
For a complete description of the method, 
see Ma\'{\i}z-Apell\'aniz (2004).
We considered TLUSTY (Lanz \& Hubeny, 2002) atmosphere models for the 
spectral energy distribution (SED) of O-type stars.
An effective temperature of $T_{\rm eff}=40000$ and $\log\,g=4.0$ 
were adopted to constrain the models. Using the six color photometry
($UBVIJK$), we derived a colour excess $E(4405-5495)=1.21\pm0.02$ 
and a ratio of total--to--selective extinction $R_{5495}=4.15\pm0.09$, 
which are the monocromatic equivalents to the usual $E(B-V)$ and $R_V$, 
respectively. 
The main source of error comes from the adopted values 
of the magnitudes, which were obtained at different epochs, and therefore 
they could correspond  to different orbital phases and be affected by the
photometric variations of the binary system.

With the bolometric magnitudes derived from our light-curve analysis, and
assuming that the bolometric corrections corresponding for the spectral types
as listed by Martins et al. (2005a) hold for the \fo\, binary components, we
obtain absolute visual magnitudes of -4.2 and -3.2 for the primary and
secondary, respectively. These values are about 1 magnitude fainter than
those tabulated by Martins et al. (2005a) for stars of spectral types
O5.5V and O9.5V, but are in closer agreement with the absolute magnitudes for
ZAMS O type stars as tabulated by Hanson et al. (1997). The values of
absolute magnitudes of the binary components that we have obtained coupled 
with the total--to--selective extinction ratio
R=4.15 derived above, would locate \fo\, at a distance of 2.2kpc. This is
coincident with the distance derived for $\eta$ Car by Davidson et al. (2001)
based on spatially resolved Doppler velocities of the bipolar ejecta.

\fo\, is located in the region of the Carina Nebula named South Pillars by
 Smith et al. (2000) (see Fig.~\ref{mapa})
recently imaged by the Spitzer Space Telescope. Newborn stars in dust pillars
 pointing to $\eta$ Car are observed in the infrared image of Spitzer 
(Smith et al. 2005).  If \fo\, is at the same distance, 
it can be considered as a ZAMS star embedded in an active star formation region.

\section{X--ray data}
As mentioned previously, \fo\ was detected as an X-ray source in EINSTEIN 
observations of the Carina Nebula. Subsequent X-ray data of \fo\ were observed 
by the \rosat\, satellite in the context of the X-Mega international campaign 
(cf. Corcoran 1996) 
which also involves spectroscopic observations of hot stars with detectable 
X-ray emission.
The \rosat\,-HRI image shows a weak X-ray emission at the position of \fo\,,
considerably weaker than expected on the basis of the first \einstein\, 
observation of this star, thus suggesting a variable X--ray source.

\begin{figure*}
\centering
\vspace{10cm}
\includegraphics{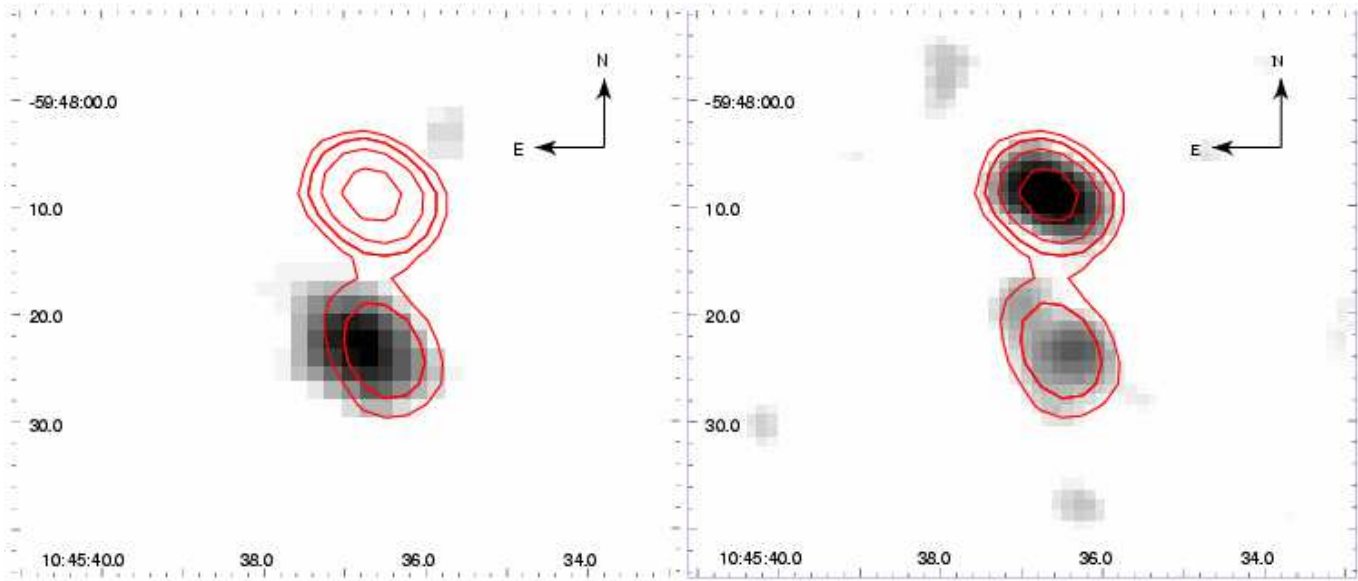}
\caption{ Portions of X-ray images  of the Carina Nebula field surrounding \fo\ 
observed by XMM (left) and Chandra (right) satellites. Contours corresponding to
0.05, 0.8, 1.5, 3. of the background counts of the Chandra image are 
superposed to the gray scale images. The brighter source to the N of \fo\, 
in the Chandra image is the source 106 in Evans et al. (2003). This source is
absent in the image obtained by XMM.
}
\label{xx}
\end{figure*}

In further X--ray observations by the \xmm\ satellite, \fo\ appeared 
as a hard X-ray source (Albacete-Colombo et al. 2003).
However, in their analysis of an X-ray image of \ec\, region obtained by the
Chandra satellite, Evans et al. (2003) did not mention \fo\, in their
list of sources with OB optical counterparts. In their list of detected sources
  without optical counterparts, the source number 106 appears close to \fo\, 
but with a difference of over 10 arcsec in the
published positional coordinates. Although \fo\, is observed near the borders
of the X-ray images, which all had \ec\, as the aimpoint, a positional
difference larger than 10 arcsecs seemed improbable.
Therefore, we decided to re-examine the Chandra X-ray image in order to see
if there might be some instrumental problem in the non-detection of \fo\, in
this image. The result is depicted in Fig.~\ref{xx}, which shows two very close
X-ray sources, the southern one is coincident with \fo\, and the northern with
source 106 of Evans et al. (2003). This source does not appear in the
X-ray image of XMM satellite used by Albacete-Colombo et al. (2003), as seen
in Fig.~\ref{xx}. Therefore, it most probably is a variable source 
due to a pre-main
sequence star located in the vicinity of \fo\,. Indeed, Evans et al. (2003) 
already have suggested that the X-ray sources without optical counterparts,
such as their source 106 close to \fo, probably are pre-main sequence stars.

As mentioned above, X-ray flux variability seem to be present 
in this system.  Albacete-Colombo \& Micela (2005)
have studied the X-ray emission of \fo\, in the 0.4-10 keV energy range,
revealing the existence of long term X-ray variability. 
Higher and lower un-absorbed X-ray flux limits are between
15.8$\times$10$^{-13}$ to 0.96$\times$10$^{-13}$ erg\,s${-1}$, implying 
L$_{\rm x}$/L$_{\rm bol}$ ranges from 9.0$\times$10$^{-7}$ to
0.54$\times$10$^{-7}$, respectively.
These authors also discuss the origin of the observed hard X-ray photons
as Inverse Compton scattering, 
and confirm the existence of soft (0.2-1.2 keV) 
short-term variability ($\sim$25\% of the total flux).

\section{Summary of the results}
\begin{itemize}
\item
We have discovered that the O-type star \fo\, immersed in the active star
formation site called 'southern pillars' in the Carina nebula, is a short 
period eclipsing binary. 

\item Both binary components are visible in the spectrum, 
the secondary component showing considerably weaker lines.
\item We classify the primary spectrum as O5.5Vz, i.e. as an early type 
Zero-Age-Main-Sequence star. The secondary seems to be of spectral type O9.5V.

\item Analysis of the ASAS light curve of \fo\, fitting a binary model by
the Wilson-Devinney method, yields an orbital inclination of $\sim$80$^\circ$.

\item The stellar masses of the components are  $\sim$ 30 and 16~$\modot$.

\item Simultaneous light and radial velocity curve analysis yields components
with smaller radii and fainter absolute magnitudes when compared with normal
galactic O-type stars.  These values are in agreement with recently born
ZAMS O type stars.

\item An individual determination of total-to-selective extinction ratio (R)
for \fo\, yields a value of 4.15, which coupled with the values of absolute
magnitudes determined from the light curve locate \fo\, at a distance of
2.2kpc, coincident with that of $\eta$ Car.

\item A Chandra X-ray image shows two close sources at the position of \fo\,.
Presumably the northern source is a pre-main-sequence star with an ocassional
high X-ray state, as this source apparently is not visible in the X-ray
image of the same location observed by the XMM satellite.
\end{itemize}
\section{acknowledgements}
We are indebted to Roberto Gamen for kindly obtaining 4 spectra of FO15
for this study.
We thank the directors and staff of CASLEO, LCO, CTIO and ESO, 
for the use of their facilities.
We also acknowledge the use at CASLEO of the CCD and data acquisition
system partly financed by U.S. NSF grant AST-90-15827 to R. M. Rich.
This research has received financial support from IALP, CONICET, Argentina.
VN and EFL thank CIC-BA for travel support. RB has received financial support
from  FONDECYT No 1050052. MO is grateful for financial support from UNLP
in the form of a research studentship.
We thank the referee, Dr. C. Evans, for useful comments which have improved
the presentation of this paper.

\label{lastpage}

\end{document}